\documentclass[preprint]{ptephy_v1}
\preprintnumber{ALBERTA-THY-7-24} 
\usepackage{hyperref}

\usepackage[a4paper, total={7in, 9in}]{geometry}
\usepackage{amsmath,graphicx,xcolor,epsfig,slashed}
\usepackage{appendix}
\usepackage{tabularx}
\usepackage{float}
\usepackage{pdfpages}
\usepackage{epstopdf}
\usepackage{caption}
\usepackage{multirow}
\usepackage{booktabs,siunitx}
\usepackage{hyperref}
\usepackage{subcaption}
\usepackage{natbib}
\usepackage{soul}

\def\vslash{v\!\!\!\slash}
\def\zslash{z\!\!\!\slash}

\def\qslash{q\!\!\!\slash}

\def\varepsilonslash{\varepsilon\!\!\!\slash}

\newcommand{\nn}{\nonumber}

\newcommand{\beq}{\begin{equation}}
\newcommand{\eeq}{\end{equation}}
\newcommand{\bqa}{\begin{eqnarray}}
\newcommand{\eqa}{\end{eqnarray}}
\newcommand{\bseq}{\begin{subequations}}
\newcommand{\eseq}{\end{subequations}}

\newcommand{\LmbdQCDD}{\Lambda_{\mathrm{QCD}}}
\newcommand{\GeV}{\mathrm{GeV}}

\begin{document}

\title{Semileptonic $W$ Decay to the $B$ Meson with Lepton Pairs in Heavy Quark Effective Theory Factorization upto $\mathcal{O}$$(\alpha_s)$}

\author{Saadi Ishaq}
\affil{School of Natural Sciences, Department of Physics, National University of Sciences and Technology, Sector H-12, Islamabad, Pakistan \email{saadiishaq@sns.nust.edu.pk}}

\author{Sajawal Zafar}
\affil{School of Natural Sciences, Department of Physics, National University of Sciences and Technology, Sector H-12, Islamabad, Pakistan \email{szafar.msphy21sns@student.nust.edu.pk}}

\author{Abdur Rehman}
\affil{Physics Department, University of Alberta, Edmonton, T6G 2E1, Alberta, Canada\email{rehman3@ualberta.ca}}

\author{Ishtiaq Ahmed}
\affil{National Centre for Physics, Quaid-i-Azam University Campus, Islamabad, Pakistan\email{ishtiaq.ahmed@ncp.edu.pk}}

\begin{abstract}%
Motivated by the study of heavy-light meson production within the framework of heavy quark effective theory (HQET) factorization, we extend the factorization formalism for a rather complicated process $W^+\to B^+\ell^+\ell^-$ in the limit of a non-zero invariant squared-mass of dilepton, $q^2$, at the lowest order in $1/m_b$ up to $\mathcal{O}(\alpha_s)$. The purpose of the current study is to extend the HQET factorization formula for the $W^+\to B^+\ell^+\ell^-$ process and subsequently compute the form factors for this channel up to next-to-leading-order corrections in $\alpha_s$. We explicitly show the amplitude of the $W^+\to B^+\ell^+\ell^-$ process can also be factorized into a convolution between the perturbatively calculable hard-scattering kernel and the non-perturbative yet universal light-cone distribution amplitude (LCDA) defined in HQET. The validity of HQET factorization depends on the assumed scale hierarchy $m_W \sim m_b \gg \Lambda_{\mathrm{QCD}}$. Within the HQET framework, we evaluate the form factors associated with the $W^+ \rightarrow B^+\ell^+\ell^-$ process, providing insights into its phenomenology. In addition, we also perform an exploratory phenomenological study on $W^+ \rightarrow B^+\ell^+\ell^-$ by employing an exponential model for the LCDAs for $B^+$ meson. Our findings reveal that the branching ratio for $W^+ \rightarrow B^+\ell^+\ell^-$ is below $10^{-10}$. Although the branching ratios are small, this channel in high luminosity LHC experiments may serve to further constraints the value of $\lambda_B$.
\end{abstract}

\subjectindex{B65 Perturbative QCD}

\maketitle

\section{Introduction}
Since the discovery of heavy-light mesons, the exclusive $B$-decays in particular not only offer an excellent laboratory to extract the SM parameters or to look for yet-unknown particles and interactions but also help to pin down the strong interaction dynamics at different scales from the Quantum Chromodynamics (QCD) point of view. Over the past few decades, 
a vast amount of literature has been devoted to 
heavy-light mesons whose underlying weak decays are understandable, but the complications appear for their theoretical elucidation in the context of perturbative and non-perturbative QCD effects.

For their theoretical description, numerous  
techniques have been introduced to disentangle perturbative and non-perturbative effects of QCD that rely on the relatively large mass of the $b$-quark as compared to the strong interaction scale $\Lambda_{\mathrm{QCD}}$. The mass of the bottom quark $m_b$ 
provides a scale at which the strong coupling $\alpha_s$ is smaller such that the short-distance effects are possible to calculate in a perturbative manner. Aiming to deal with non-perturbative effects, various theoretical approaches are developed. Among them, 
the QCD factorization has emerged as the predominant theoretical framework, which derives from the first principle~\cite{Beneke:1999br,Beneke:2000ry,Beneke:2001ev}.

The most straightforward application of QCD factorization is for the exclusive heavy-to-light radiative $B$ meson transitions~\cite{Korchemsky:1999qb,Lunghi:2002ju,DescotesGenon:2002mw}. Where the amplitude can be factorized, in the heavy quark limit, as a convolution between hard-scattering kernel perturbative in nature and non-perturbative light-cone distribution amplitudes (LCDAs) of $B$-mesons.
Furthermore, the Lorentz invariant form factors of the exclusive $B$-decays can be calculated in terms of process-independent $B$ mesons LCDAs. As a result, the radiative $B$-decays serve as an excellent testing ground for exploring the characteristics of $B$-meson LCDAs and verifying the validity of the factorization approach.
This factorization formula has also been extended for somewhat more complicated channels, e.g., $B\rightarrow M_1 M_2$~\cite{Beneke:2000ry,Beneke:2001ev} and $B\rightarrow \gamma\gamma$ \cite{Bosch:2002bv}.

The significant generation of weak gauge bosons at the Large Hadron Collider (LHC) is a source of inspiration to validate the predictions of the SM, search for new physics (NP), improve our understanding of QCD dynamics at different regimes, and also offer opportunities to investigate the exclusive $W$ gauge boson decays. Among these decay modes, $W\rightarrow D_{s}\gamma$ stands out with the highest branching fraction. The first detailed analysis of the radiative decay process $W\rightarrow D_{s}\gamma$ was studied decades ago \cite{Arnellos:1981gy,Keum:1993eb}. The upper limit was set by CDF collaboration with the value $\text{Br}^{\text{Exp}}(W\rightarrow D_s \gamma) < 1.3\times 10^{-3}$~\cite{Abe:1998vm}. This high-yield production of $W^{\pm}$ and $Z$ has been a pivotal driver in unravelling their decay characteristics with increased precision. In this context, several exclusive radiative decays of $W$ and $Z$ bosons into heavy-light meson have been investigated in the standard collinear factorization (or light-cone factorization)~\cite{Grossmann:2015lea}.
The heavy-meson LCDAs that appear in this factorization formula are not completely non-perturbative, as they still entail the hard scale because the $b$-quark field is defined in full QCD.

Nonetheless, the LCDA of heavy-light meson entering the HQET factorization formula is entirely nonperturbative because the $b$-quark field is defined in HQET rather than in full QCD. It is noteworthy to highlight that both types of LCDA associated with heavy-light meson are connected through a perturbatively calculable matching coefficient~\cite{Ishaq:2019dst}. The HQET factorization formula crucially depends on the mass hierarchy: $m_W \sim m_b \gg \Lambda_{\mathrm{QCD}}$. This hierarchy ensures that the LCDA's dependence is confined to the soft scale; consequently, the LCDA's behaviour is not entangled with perturbative effects. This separation of scales, facilitated by the mass hierarchy, allows for a more tractable and precise description of the heavy meson HQET LCDA. Notably, the production of heavy-light mesons within the HQET factorization formalism has been comprehensively addressed in~\cite{Ishaq:2019zki} up to NLO in $\alpha_s$, shedding light on the application of this factorization formula in the study of these processes. Furthermore, the exclusive production of flavoured quarkonia, such as $W^+ \rightarrow B_c^+ \gamma$, has also been investigated through $\mathcal{O}(\alpha_s)$ within the NRQCD factorization framework~\cite{Feng:2019meh}.

The HQET factorization formula for the decay process $W^+ \rightarrow B^+\gamma$ has previously been established in~\cite{Ishaq:2019zki} for the scenario where photon is energetic,
$q^-\gg \Lambda_{\mathrm{QCD}}$. This factorization formula for the production $B$ mesons contains only ${\phi}^{+}_{B}(\omega)$, the leading-twist $B$ meson LCDA. Consequently, the relevant transition form factors expressed in terms of poorly constraint first-inverse moment $1/ \lambda_B\equiv\int_0^\infty d\omega ~\phi_B^+(\omega)/ \omega $, in the same fashion as appeared in QCD calculation of exclusive $B$ meson decays. To constraint $\lambda_B$, the exclusive production of $B$ meson through $W$ radiative decay and the radiative $B$-decays would serve as clean channels. One can find the most recent constraint on $\lambda_B$ from BELLE~\cite{Belle:2018jqd}, this result can be anticipated to be updated at BELLE II. However, implementing this strategy at LHCb is challenging due to the difficulty in reconstructing the photon involved in the radiative $B\rightarrow \gamma\ell\nu$ decay. The analysis would be feasible once the photon further decays into dileptons. Following the measurement reported by LHCb experiment~\cite{LHCb:2018jvy} on branching ratio, $\text{Br}(B^+\rightarrow \mu^+\bar{\nu}_{\mu}\mu^-\mu^+)< 1.6\times 10^{-8}$, recent theoretical studies have been focused on $B$-decays to four-leptons~\cite{Bharucha:2021zay,Beneke:2021rjf} within the QCD factorization framework to constraint the phenomenological parameter $\lambda_B$. 

In light of these developments, there is a compelling interest in investigating the potential of HQET factorization for a complicated process $W^+\rightarrow B^+\ell^+\ell^-$. In this paper, our study extends the HQET factorization formula for the process $W^+\rightarrow B^+\ell^+\ell^-$ in which $q^2$ is non-zero, where $q^2$ is the invariant mass squared of the $\ell^+\ell^-$ pair originating from the virtual photon. Our primary objective is to perform a comprehensive calculation of the form factors associated with the $W^+ \rightarrow B^+\ell^+\ell^-$ process for both intermediate $q^2\sim\mathcal{O}(m_b\,\LmbdQCDD)$ and small scale $q^2 << m_B^2$, within HQET factorization framework, up to NLO in $\alpha_s$ at lowest order in $1/m_b$. We also explore the capability of $W^+ \rightarrow B^+\ell^+\ell^-$ to constraint the $\lambda_B$ and hence could provide an alternative measurement. 

The article is organiszd as follows: In the next section, we provide the necessary definitions and notations to specify the setup for the calculation of the hard scattering kernel in HQET. In Section ~\ref{sec:hrd_scttr_amp}, we present the details of perturbative QCD calculation on the hard scattering kernel of the $W^+\rightarrow B^+\ell^+ \ell^-$ decay process. In Section ~\ref{sec:phnmnlgy}, we study the phenomenology of this decay process and report numerical predictions on the branching ratio of $W^+\rightarrow B^+\ell^+ \ell^-$ by invoking an exponential LCDA model. We summarise in Section~\ref{sec:sumry}.
\section{Definitions and Notations}\label{sec:def_not}
In this section, we introduce the notations and definitions used in this work. We start with the kinematics for the process $W^+\rightarrow B^+\ell^+\ell^-$ where the momentum of $W^+$ is represented by $Q$ with $Q=P+q$. Here $P$ is the momenta of $B^+$ and $q\,(=q_1+q_2)$ is the momenta of dilepton $(\ell^+\ell^-)$ satisfying $q^2\neq 0$. The polarization vector for the $W^+$ is denoted by $\varepsilon_{W}$.

For convenience, we work in light-cone coordinates by introducing light-like reference vectors,
$n_{\pm}^{\mu}\equiv\frac{1}{\sqrt{2}}(1,0,0,\mp 1)$ that satisfy the conditions: $n_\pm^2=0$ and $n_+\cdot n_-=1$. It allows to write any four-vector $a^{\mu}=(a^0,a^1,a^2,a^3)$ as
\beq
a^{\mu}=(n_{-}\cdot a)
n_{+}^{\mu}+ (n_{+}\cdot a)n_{-}^{\mu}+a^{\mu}_{\perp}\equiv
a^+ n_{+}^{\mu}+a^- n_{-}^{\mu} + a^{\mu}_{\perp},
\eeq
where ${a}^{\mu}_{\perp}=(0,a^{1},a^{2},0)$ represents the transverse component of the four vector.
It would be convenient to stick with the $W$ boson rest frame to investigate this process as long as we follow the standard collinear factorization approach. In the frame where the $W$-boson is at rest, we assume that the $B$ meson moves along the positive $\hat{z}$ axis,
while the virtual photon moves in the opposite direction to the $B$ meson.

Since the HQET is formulated in the rest frame of $B$ meson, therefore, it is inevitable to boost this process to the $B$ meson rest frame. To achieve this, a dimensionless four velocity $v^\mu$ is introduced via $P^{\mu}=m_{B} v^{\mu}$, satisfying $v^{2}=1$. The momentum of the virtual photon in the light-cone basis can be decomposed as
\beq
q^{\mu}=(n_{-}\cdot q)
n_{+}^{\mu}+ (n_{+}\cdot q)n_{-}^{\mu}\equiv
q^+ n_{+}^{\mu}+q^- n_{-}^{\mu}\, .
\eeq
The large component of $q^{\mu}$ reads as
\beq
q^- = \frac{1}{\sqrt{2}}\left(\frac{m_{W}^{2}-q^2 -m_{B}^{2}+\sqrt{\lambda}}{2m_{B}}\right)\,,
\eeq
where $\lambda\equiv m_{W}^{4}+(m_{B}^2-q^{2})^2-2m_{W}^{2}(q^{2}+m_{B}^2)$. The physical range of the invariant squared-mass of a dilepton is $4 m_{\ell}^{2} \le q^2\le (m_W-m_B)^2$. It is crucial to identify that the amplitude for the exclusive production of a heavy-light meson is highly suppressed for very large $q^2\sim\mathcal{O}(m_B^2)$ at the heavy quark limit. However, an interesting scenario emerges when $q^2 << m_B^2$, while one component of $q^\mu$ still remains large, $q^-\sim\mathcal{O}(m_B)$, and $q^+$ is an order of $\LmbdQCDD$ or even smaller. Hence, with the aid of $q^2=2 q^+ q^-\equiv s$, one can deduce that the ``+''-component of $q^{\mu}$ is suppressed relative to $q^-$.

For our convenience, the transition amplitude of exclusive $W^+\to B^+\ell^+\ell^-$ decay can be written as
\begin{equation}
\mathcal{M}=q^2~\mathcal{\Tilde{M}}(W^+\to B^+\ell^+\ell^-)\,,
\label{Ampl:Lorentz:decompSimp}
\end{equation}
where
\begin{eqnarray}
 \mathcal{\Tilde{M}}&=&\frac{e_u e^3 V_{ub}}{4 q^{2}\sqrt{2}\text{sin}\theta_W}\bigg[\varepsilon_{\mu\nu\alpha\beta}\frac{P^\nu q^\alpha\varepsilon^\beta_W}{P\cdot q}F_V
 +i\left(\varepsilon_{W\mu}+\frac{P_\mu q\cdot\varepsilon_W}{P\cdot q}\right)F_A\bigg]\bar \ell\gamma^\mu \ell,
\label{Ampl:Lorentz:decomp}
\end{eqnarray}
where $e$ and $e_{u}$ are the electric charges of leptons and $u$ quark, respectively, $\theta_W$ is the weak mixing angle, and $V_{ub}$ denotes the CKM matrix element. The scalar quantities,
$F_{V}$ and $F_A$ are the vector and axial-vector form factors, respectively, for the process $W^+\to B^+\ell^+\ell^-$. These form factors depend on the kinematical variables, $m_W$, $m_B$, which encode the non-trivial QCD dynamics and have to be computed for the predictions of exclusive heavy-light meson production. The decay rate for the processes $W^+\to B^+\ell^+\ell^-$ in the $W$ rest frame reads as
\begin{eqnarray}
    \frac{d\Gamma}{dsdt}=\frac{1}{256\pi^3m_W^3 q^4}|\mathcal{M}|^2\,, \label{DR}
\end{eqnarray}
where
\begin{eqnarray}
    &&|\mathcal{M}|^2=\frac{\big(|F_V|^2+|F_A|^2\big) |V_{ub}|^2}{m_W^2
   \left(m_B^2-m_W^2+s\right)^2}\times\bigg(\frac{2 e_u^2\pi ^3 \alpha^3} {\text{sin}^2\theta_W}\bigg)\bigg[4 m_\ell^4 s \left(m_B^2+m_W^2\right)\notag \\
   &&+2m_\ell^2 \left(m_B^6+m_B^4 \left(m_W^2+3 s\right)-m_B^2 \left(5m_W^4-2 m_W^2 s+s (3 s+4 t)\right)+3 m_W^6-5 m_W^4
   s\right)\notag \\
   &&+2
   m_\ell^2 \left(m_W^2 s (3 s-4 t)-s^3\right)+2 m_B^8-m_B^6 \left(6m_W^2+s+2 t\right)+2 t \left(m_W^6-3
   m_W^4 s+m_W^2 s^2+s^3\right)\notag \\
   &&+m_B^4 \left(6 m_W^4+m_W^2 (9 s+6 t)-3 s(s+2 t)\right)+4 m_W^2 s t^2+s
   \left(m_W^2-s\right)^2 \left(3 m_W^2+s\right)\notag \\
   &&+m_B^2 \left(-2 m_W^6-3 m_W^4 (s+2 t)-4 m_W^2
   s t+s \left(s^2+6 s t+4 t^2\right)\right)\bigg],\label{DR1}
\end{eqnarray}
with $\alpha$ is the QED fine structure constant and
\vspace{-0.45cm}
\begin{eqnarray}
s_{min}&=&4m_{\ell}^2\, ;\qquad\qquad s_{max}=(m_W-m_B)^2,\notag \\
t_{max(min)}&=&\frac{1}{2} \left(2
   m_{\ell}^2+m_B^2+m_W^2-s\pm\sqrt{\lambda } \sqrt{1-\frac{4 m_{\ell}^2}{s}}\right),\notag\\
   \lambda &=&m_B^4-2 m_B^2 m_W^2-2 m_B^2 s+m_W^4-2
   m_W^2 s+s \,.
\end{eqnarray}
In the following sections, we compute the analytical expression of the form factors $F_{V/A}$ up to NLO QCD correction at leading order in $1/m_b$. Note that any frame of reference can be used to compute these form factors because they are Lorentz scalar. In this study, we stick with the $B$ rest frame for their computation to make the picture of HQET factorization more transparent.
\subsection{B meson LCDA in HQET}\label{sec:HQET LCDA}
In this subsection, we summarize the pivotal aspects of $B$ meson LCDA. Within the framework of factorization, $B$ meson LCDA emerges as the main nonperturbative input for describing the numerous exclusive decays and production processes involving $B$ meson. Since the LCDA of heavy-light mesons can be defined either in standard QCD or HQET, however, the convolution of the hard scattering kernel with the LCDA remains identical. In the HQET factorization approach, the LCDA of $B$ meson can be expressed as an independent pair of nonperturbative functions $\widetilde{\phi}_{B}^{\pm}$~\cite{Grozin:1996pq,Beneke:2000wa}
\beq
\langle B(v)|\bar
{u}_{\beta}(z)[z,0] h_{v,\alpha}(0)|0\rangle=\frac{i \hat
f_{B}m_{B}}{4}\left\{\left[2
\widetilde{\phi}_{B}^{+}(t)-\frac{\zslash}{t}
\Big(\widetilde{\phi}_{B}^{-}(t)-\widetilde{\phi}_{B}^{+}(t)\Big)\right]
\frac{1-\vslash}{2}\gamma_{5}\right\}_{\alpha\beta},
\label{eq:LCDA_twist_decomp}
\eeq
In Eq. (\ref{eq:LCDA_twist_decomp}), the $B$ meson has been deliberately placed in the bra rather than the ket. This choice is made due to our focus on the production of $B$ meson instead of its decay processes, where $z^2 =0$, $t=v\cdot z$, and $u$ is the standard
QCD light quark field and $h_v$ refer to the $\bar{b}$ quark field
with the velocity label $v$ defined in HQET. And, $\hat{f}_{B}$ signifies the $B$ meson decay constant defined in HQET. The  $\alpha$, $\beta$ are spinor indices, and $[z, 0]$ is a light-like gauge link to ensure the gauge invariance of the LCDA,
\beq
[z, 0]=\mathcal{P} \exp \left[-i g_{s} \int_{0}^{z} d
\xi^{\mu} A_\mu^{a}(\xi) t^{a}\right],
\eeq
where $\mathcal{P}$ indicates the path ordering, $g_s$ is a strong coupling constant, and $t^{a}(a=1, \cdots, 8)$ refers to $SU(3)$ generators in fundamental representation. $A_\mu^a$ is the gauge field, and $\xi$ is the momentum distribution along the Wilson line. The QCD decay constant $f_B$ can be found throughperturbative matching~\cite{Eichten:1989zv,Neubert:1993mb},
\beq
f_B = \hat{f}_B (\mu_F) \left[ 1 - {\alpha_s C_F \over 4 \pi} \left(3\ln \frac { \mu_F}{m_b} + 2
\right) \right] +\mathcal{O}\left(\alpha_s^2\right),
\label{hat:f:matching:fB}
\eeq
where $\mu_F$ is the factorization scale and $C_F$ is the color factor. We can obtain the momentum space representation of $B$ meson LCDA by Fourier transforming the coordinate-space correlators provided in Eq. (\ref{eq:LCDA_twist_decomp}),
\beq
\Phi_{B}^{\pm}(\omega) \equiv i\hat{f}_{B}m_{B}\phi^\pm_{B}(\omega)=\frac{1}{v^{\pm}}\displaystyle\int
{dt\over 2\pi}\, e^{i\omega t}\langle B(v)|\bar
{u}(z)[z,0]\slashed n_\mp \gamma_5 h_{v}(0)|0\rangle\Big|_{z^{+},z^{\perp}=0}\,.
\label{eq:LCDA_HQET}
\eeq
Here $\omega$ indicates the ``+''-momentum carried by the spectator
quark in the $B$ rest frame, whose typical value is $\sim \Lambda_{\rm QCD}$. A pair of independent, nonperturbative functions is defined through
\begin{equation}
{\phi}^{\pm}_{B}(\omega)=\int_{0}^{\infty} {d t\over 2\pi} e^{i
\omega t} \widetilde\phi^{\pm}_{B}(t).
\end{equation}
We will see explicitly that only $\phi^{+}_{B}(\omega)$ survives in the HQET factorization approach, which contributes to the form factors $F_{V, A}$. Similar to $B$ meson LCDA defined in standard QCD, $\phi^{+}_{B}(\omega)$ is also scale-dependent, whose scale dependence is governed by Lange and Neubert~\cite{Lange:2003ff}
\begin{align}
&\notag \frac { d } { d \ln \mu } \phi _ { B } ^ { + } ( \omega ,\mu ) =\notag\\
& - \frac { \alpha _ { s } C _ { F } } { 4 \pi } \int _ { 0 } ^ { \infty } d \omega ^ { \prime }\left\{\left( 4 \ln \frac { \mu} { \omega } - 2 \right) \delta \left( \omega - \omega ^ { \prime } \right) - 4 \omega \left[ \frac { \theta \left( \omega ^ { \prime } - \omega \right) } { \omega^{ \prime } \left( \omega ^ { \prime } - \omega \right) }\right.\right.
\left.\left.+ \frac { \theta \left( \omega - \omega ^ { \prime } \right) }
{ \omega \left( \omega - \omega ^ { \prime } \right) } \right] _ { + }\right\}\phi_{ B }^{+}
\left( \omega ^ { \prime } , \mu \right),
\nn\\
\label{LN:evolution:eq}
\end{align}
where $\mu$ is the renormalization scale. $\phi^{+}_{B}(\omega)$ can be deduced at some initial scale $\mu_0=1\,\GeV$,
one can then determine its form at any other scale (typically, $1\,\GeV\le\mu\le m_b$) by
solving the evolution equation~\eqref{LN:evolution:eq}. Whereas, the scale dependence of the meson LCDA defined in full QCD is governed by the Efremov-Radyushkin-Brodsky-Lepage (ERBL) equation~\cite{earlyBL1, earlyBL2, earlyBL3, earlyBL4, Efremov:1979qk,earlyCZ1,earlyCZ2,earlyCZ3}.

\section{HQET factorization and Form factors for $W\to B ~ \ell^+\ell^-$}\label{sec:hrd_scttr_amp}
\label{sec:HQET:factorization}
This section is devoted to present the calculation of the hard scattering kernel at $\mathcal{O}(\alpha_s)$ to the leading order in $1/m_b$. The computation of the form factors $F_{V, A}$ is conducted in the $B$ meson's rest frame. For this, we follow the NLO calculation of $W^+\rightarrow B^+ \gamma$ ~\cite{Ishaq:2019zki} and therefore, it is instructive to modify the HQET factorization formula reported in~\cite{Ishaq:2019zki} for $W^+\rightarrow B^+ ~ \ell^+\ell^-$ as
\beq
{\mathcal M}=e~\bar \ell\gamma^\mu \ell \int_0^\infty \!\! d\omega\, T_{\mu}(\omega, m_b,q^2, \mu_F)
\Phi^+_B(\omega,\mu_F)+\mathcal{O}\left(m_b^{-1}\right),
\label{HQET:factorization:theorem}
\eeq
where $T_{\mu}(\omega,m_b, q^2,\mu_F)$ is the hard-scattering kernel, which can be computed in perturbation theory by employing the perturbative matching technique. The hard scattering kernel is also a function of the invariant squared mass of the dilepton, $q^2$. In the following section, it is explicitly shown that the hard-scattering kernel for the $W\to B+\gamma$ process can be recovered by the substitution of $q^2 \rightarrow 0$ in the expression of $T_{\mu}(\omega,m_b, q^2,\mu_F)$. In perturbative calculation, the hard kernel is independent of the external state, and we safely choose a convenient partonic state as $B^+$ meson consists of a $\bar{b}-$antiquark and a $u$-quark. As a result, the LCDA in Eq.~\eqref{eq:LCDA_HQET} takes the form
\beq
\Phi_{[\bar b u]}^{\pm}(\omega)=\frac{1}{v^{\pm}}\displaystyle\int {dt\over 2\pi}\, e^{i\omega t}\langle [\bar b u](P)|\bar {u}(z)[z,0]\slashed n_\mp \gamma_5 h_{v}(0)|0\rangle\Big|_{z^{+},z^{\perp}=0}.
\label{eq:LCDA_bbaru}
\eeq
To extract the amplitude of the $W^+\to B^+\ell^+\ell^-$ process in the factorization approach, one can use the $B$ meson momentum projector. For this, the substitution reported in~\cite{Beneke:2000wa} may be incorporated at the quark-level amplitude
\begin{equation}
v_i (P\!-\!k) \bar{u}_j (k)\to {\delta_{ij}\over N_c} \frac{i \hat{f}_{B}
m_b}{4}\!\left\{\frac{1\!-\!\slashed v}{2}\!\left[\phi_{B}^{+}(\omega)
{\slashed n_{+}\over\sqrt{2}}\!+\!\phi_{B}^{-}(\omega) {\slashed
n_{-}\over \sqrt{2}}-\omega \phi_{B}^{-}(\omega) \gamma^{\mu}_{\perp}
\frac{\partial}{\partial k_{\perp\mu}}\right]\!\!
\gamma_{5}\right\}\Bigg|_{k=\omega v}.
\label{HQR:spin:projector:B}
\end{equation}
where $i,j=1,2, \cdots, N_c$ are color indices and $N_c=3$.
The first Kronecker symbol serves the color-singlet projector. The momentum of spectator quark $u$ is $k^{\mu}$, which is soft and scales as $k^{\mu}\sim \LmbdQCDD$.

The HQET factorization approach for the exclusive production of heavy-light mesons arises from the heavy-quark recombination (HQR) mechanism~\cite{Braaten:2001bf}, specifically for the color-singlet channel. This mechanism achieved noteworthy success in explaining the observed charm/anticharm hadron production asymmetry~\cite{Braaten:2001uu,Braaten:2002yt,Braaten:2003vy}. The HQR mechanism offers a shortcut to efficiently replicate the heavy meson production amplitude with less computational effort, in contrast to ~\cite{Beneke:2000wa}. For this, one can invoke the following projector on the quark-level amplitude
\beq
v_i (P-k) \bar{u}_j (k)\to  {\delta_{ij} \over N_c}
 \frac{1-\slashed v}{4} \gamma_5\,.
\label{eq:projector:bbaru}
\eeq
which ensures that the fictitious $B^+$ meson is the color and spin-singlet. It is noteworthy that this simplification has been utilized to evaluate the NLO correction to form factors for the $W\to B+\gamma$ process~\cite{Ishaq:2019zki} in HQET factorization and for $W\to B_c+\gamma$ within the NRQCD factorization framework\cite{Jia:2010fw,Feng:2019meh}.
\subsection{Factorization at Tree level}
\label{sec:tree:level:hard:kernel}
\begin{figure}
\centering
\includegraphics[clip,width=0.8\textwidth]{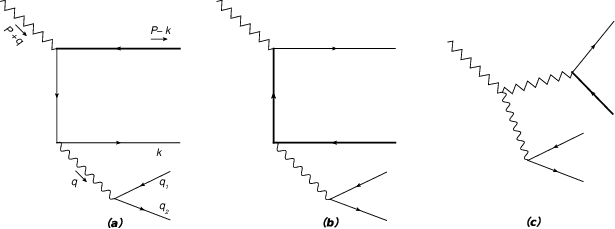}
\caption{The Feynman diagrams for $W^{+}\rightarrow [\bar{b} u]+\ell^+\ell^-$ at tree level. The bold line represents the $\bar b$ quark.}
\label{fig:tree}
\end{figure}
At the tree level, only three diagrams contribute to the quark-level process, $W\to [\bar{b}(P-k) u(k)]+\ell^+\ell^-$, as shown
in Fig.(\ref{fig:tree}).
The momentum of $u$-quark  scales as $\left(k^+,k^-,|\boldsymbol{k}_\perp|\right)
\sim\mathcal{O}(\Lambda_{\mathrm{QCD}},\Lambda_{\mathrm{QCD}},\Lambda_{\mathrm{QCD}})$ while the momenta of dilepton scale as $\left(q^+,q^-,|\boldsymbol{q}_\perp|\right)
\sim\mathcal{O}(\Lambda_{\mathrm{QCD}},m_{b},0)$. Therefore, it is easy to identify that the $u$-quark propagator in Fig.(\ref{fig:tree}a) is of $\mathcal{O}(1/\LmbdQCDD)$ which is leading order, while in Fig.(\ref{fig:tree}b) the internal propagator is
$\mathcal{O}(1/m_{b})$. Thus, the Fig.(\ref{fig:tree}b) is suppressed by one power of $m_{b}$ compared to Fig.(\ref{fig:tree}a), consequently, the contribution of this diagram could be neglected at the leading twist. Fig.(\ref{fig:tree}c) is also dropped because it is at $\mathcal{O}\left(1/m_W^{2}\right)$ due to the intermediate $W$-propagator. Hence the LO contribution comes from the diagram in which the virtual photon is radiated from the spectator quark. In the light of \eqref{Ampl:Lorentz:decompSimp} the tree-level QCD amplitude
in the heavy quark limit reads as:
\begin{equation}
\mathcal{M}^{(0)}=q^2~\mathcal{\Tilde{M}}^{(0)}(W^+\to B^+\ell^+\ell^-)\,,
\end{equation}
where $q^2$ represents the propagator of virtual photon and $\mathcal{M}^{(0)}$ takes the form
\bqa
\mathcal{M}^{(0)}
&=&\frac{e_{u} e^2 V_{ub} }{2\sqrt{2}\sin\theta_W (q^2+2q^-
k^+)}\bar{u}_j (k) \gamma_{\mu}\qslash\varepsilonslash_W(1-\gamma_{5}) v_i (P\!-\!k) \big( e~\bar \ell\gamma^\mu \ell \big),
\label{eq:M_LO}
\eqa
\bqa
\mathcal{M}^{(0)} &\approx & {\frac{e_{u} e^2 V_{ub} }{4\sqrt{2}\sin\theta_W (q^+ q^-+q^-
k^+)}}\text{Tr}\Big[ \frac{1-\slashed v}{4} \gamma_5 \gamma_{\mu}\qslash\varepsilonslash_W(1-\gamma_{5})\Big]\big( e~\bar \ell\gamma^\mu \ell \big),
\nn\\
\nn\\
 &=& \frac{e_{u}e^{2}V_{ub}}{4\sqrt{2}\sin\theta_{W}}\left( -i\frac{\epsilon_{\mu\nu\alpha\beta} v^{\nu}\epsilon^{\alpha}n_{-}^{\beta}}{v^{+}}+\epsilon_{\mu}-\frac{v_{\mu}\epsilon_{\beta}n_{-}^{\beta}}{v^{+}}\right)\big( e~\bar \ell\gamma^\mu \ell \big)\int\frac{d \omega}{\left(\omega+ q^{+}/v^{+}\right)}\delta\left(k^{+}/v^{+}-\omega\right).\nn\\
\label{eq:M_LO1}
\eqa
%
Whereas, the LCDAs  for the fictitious $B^+$ meson given in Eq.  \eqref{eq:LCDA_bbaru} come out in the following simple form
%
\beq
\Phi_{[\bar b u]}^{\pm (0)}(\omega)=\frac{1}{v^{\pm}}\delta\left(k^{+}/v^{+}-\omega\right)\text{Tr}\Big[ \frac{1-\slashed v}{4} \gamma_5 \slashed n_\mp\gamma_5\Big],
\label{eq:LCDA_tree1}
\eeq
%
and found to be
\beq
\Phi_{[\bar b u]}^{\pm (0)}(\omega)=\delta\left(k^{+}/v^{+}-\omega\right).
\label{eq:LCDA_tree}
\eeq
While the hard-kernel at tree level will be
\beq 
T_{\mu}^{(0)}=\frac{e_{u}e^{2}V_{ub}}{4\sqrt{2}\sin\theta_{W}}\left(-\textit{i}\frac{\epsilon_{\mu\nu\alpha\beta}p^{\nu}\epsilon^{\alpha}q^{\beta}}{p.q}+\epsilon_{\mu}+\frac{q.\epsilon_{W}}{p.q} p_{\mu}\right)\left(\frac{1}{\omega+ q^{+}/v^{+}}\right)\,.
\eeq
The analytical expression for form factors can be obtained by comparing the Lorentz decomposition specified in \eqref{Ampl:Lorentz:decomp} with the factorization formula proposed in Eq.\eqref{HQET:factorization:theorem}. At tree level, they read
\begin{align}
F_V^{(0)}=F_A^{(0)}&={f_B~m_B} \int_0^\infty \frac{d\omega}{\omega+ q^{+}/v^{+}} \phi_B^+\big(\omega\big)
={\frac{f_B~m_B}{\lambda^{+}_B(q^{+})}}.
\label{FVA:LO:expression}
\end{align}
By invoking Eq.\eqref{HQR:spin:projector:B}, one can prove that $\phi_B^-\big(\omega\big)$ terms vanish, and as a result, $\Phi_{[\bar{b}u]}^-$ does not enter the HQET factorization formula. Here $\lambda^{-1}_B(q^{+})$ is the inverse moment of the $B$ meson LCDA which depends on the invariant squared-mass of the dilepton through $q^+=q^2/2q^-$ and is defined as
\begin{align}
\frac{1}{\lambda^{+}_B(q^{+})}\equiv\int_0^\infty \frac{d\omega}{\omega+ q^{+}/v^{+}} \phi_B^+\big(\omega\big).
\label{Invrs:moment:LCDA}
\end{align}
Note that $\lambda^{-1}_B$ scales as $\lambda^{-1}_B\sim \LmbdQCDD$ and is also scale-dependent. For $q^2\rightarrow 0$, the expressions of form factors and inverse moment reduce to the one \cite{Ishaq:2019zki} for real photon.

\subsection{Factorization at One-loop level}
\label{sec:NLO:alphas:hard:kernel}
\begin{figure}
\centering
\includegraphics[clip,width=0.85\textwidth]{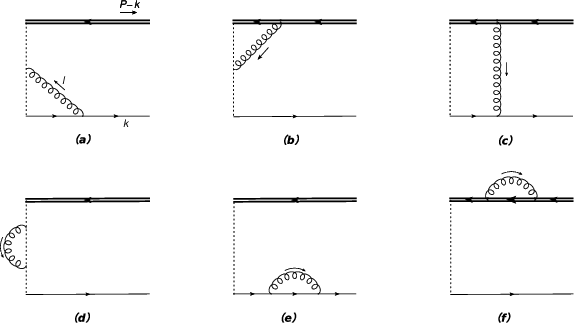}
\caption{One-loop QCD correction to LCDA for a fictitious $B$ meson.
The double line represents the $\bar{b}$ field in HQET, dashed line represents the gauge link.}
\label{Fig:One:loop:LCDA}
\end{figure}
\begin{figure}
\centering
\includegraphics[clip,width=0.8\textwidth]{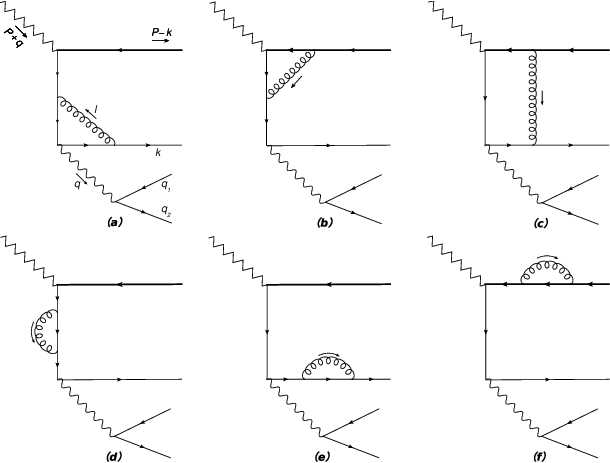}
\caption{One-loop QCD correction to the amplitude for $W^{+}\rightarrow
[\bar{b}u]+\ell^+\ell^-$. We consider only those diagrams in which the virtual photon emitted from the spectator $u$ quark.}
\label{Fig:One:loop:Amplitude}
\end{figure}

In this subsection, we present the analytical expression of the hard-scattering kernel at next-to-leading order (NLO) in $\alpha_s$. We can expand the matrix element on the left side of Eq. \eqref{HQET:factorization:theorem} in perturbation theory. Thus, up to ${\cal O}(\alpha_s)$, it takes the following schematic form
\bqa
\mathcal{M} &=& \mathcal{M}^{(0)}+ \mathcal{M}^{(1)}+{\cal O}(\alpha_s^2),
\\
&=& e~\bar \ell\gamma^\mu \ell\Big[\Big(\Phi^{(0)}\otimes T_{\mu}^{(0)}\Big)+\Big(\Phi^{(0)}\otimes T_{\mu}^{(1)} + \Phi^{(1)}\otimes T_{\mu}^{(0)}\Big)+{\cal O}(\alpha_s^2)\Big],
\label{Expanding:fact:ampl:NLO:alphas}
\eqa
where $\otimes$ encodes the convolution integral in $\omega$ and superscripts represent the power of $\alpha_s$. We calculate the $T_{\mu}^{(1)}$ by taking $B^+ =[\bar{b}(P-k) u(k)]$ as
\beq
\big( e~\bar \ell\gamma^\mu \ell \big) \big( \Phi^{(0)}\otimes T_{\mu}^{(1)}\big)=\mathcal{M}^{(1)}-\big( e~\bar \ell\gamma^\mu \ell \big)\big(\Phi^{(1)}\otimes T_{\mu}^{(0)}\big).
\label{Extracting:T1:General}
\eeq
At lower order in $1/m_b$, the dominated contribution is an order of $\Lambda_{\rm QCD}^{-1}$ that arises only through those diagrams for which the virtual photon is emitted from the spectator quark. To have the hard-scattering kernel at NLO, one needs to evaluate $\mathcal{M}^{(1)}$ from Eq. \eqref{Extracting:T1:General} in standard QCD and $\Phi^{(1)}$ in HQET. The one-loop diagrams for $\Phi^{(1)}$ and $\mathcal{M}^{(1)}$ are represented in Fig.~\eqref{Fig:One:loop:LCDA} and Fig.~\eqref{Fig:One:loop:Amplitude}, respectively. The general principle of effective field theory dictates that the infrared (IR) finite hard kernel at NLO precision can be extracted by evaluating the difference of $\mathcal{M}^{(1)}$ and $\Phi^{(1)}\otimes T_{\mu}^{(0)}$ on a diagram-by-diagram basis, as these quantities contain the same IR singularities.
Therefore, it appears instructive to regulate mass (collinear) singularity in the same way for both $\mathcal{M}^{(1)}$ and $\Phi^{(1)}\otimes T^{(0)}$, this can be achieved by taking a nonzero mass $m_u$ to the spectator $u$ quark. However, dimensional regularization (with spacetime dimensions $d=4-2 \epsilon$) is used to regularize UV divergences, and we used the $\overline{\rm MS}$ renormalization scheme by redefining the 't Hooft unit mass through $\mu^2\to \mu^2 {e^{-\gamma_E}\over 4\pi}$. The 't Hooft unit mass $\mu_R$ is designated for the QCD amplitude $\mathcal{M}^{(1)}$ calculation, while a different 't Hooft unit mass $\mu_F$ is used in computing $\Phi^{(1)}$. We perform the calculation in the Feynman gauge for our convenience.

To find out the $T_{\mu}^{(1)}(\omega)$ in the light of Eq.~\eqref{Extracting:T1:General}, we must calculate $\mathcal{M}^{(1)}$. This can be done by evaluating the one-loop QCD diagrams in Fig.~\eqref{Fig:One:loop:Amplitude}. It is easy to evaluate the electromagnetic vertex correction, weak vertex correction, and internal quark self-energy QCD diagrams as represented in Fig.~\eqref{Fig:One:loop:Amplitude}$(a)$, $(b)$ and $(d)$, respectively. We have also tacitly included those quark mass counterterm diagrams to obtain UV-finite results. On the other hand, the NLO perturbative contributions to the $\Phi^{(1)}\otimes T^{(0)}$ can be obtained from the {\it soft} loop region of the electromagnetic vertex correction, weak vertex correction, and light quark propagator correction in their respective QCD counterparts, as depicted in Fig.~\eqref{Fig:One:loop:Amplitude}. For the calculation of $\Phi_{+}^{(1)}$, one needs the Feynman rules for the Wilson line. If $p$ be the momentum flowing in the gauge link~\cite{Collins:2011zzd}, then the Feynman rules for propagator are $1/p^+$, and $-ig_sT^a n_+^\mu$ for eikonal vertex.
Applying these Feynman rules to evaluate the Fig.~\eqref{Fig:One:loop:LCDA} diagram by diagram, we found that the contribution to $\Phi_{+}^{(1)}\otimes T^{(0)}$ is the same as reported in \cite{Ishaq:2019zki}. We also noticed that there is no contribution to $\Phi_{+}^{(1)}\otimes T_{\mu}^{(0)}$ from the gauge link self-energy diagram as shown in Fig.~\eqref{Fig:One:loop:LCDA}$(d)$ because its contribution to $\Phi_{+}^{(1)}$ is proportional to $n^2_+=0$. Hence, on subtracting the contributions of $\Phi_{+}^{(1)}\otimes T_{\mu}^{(0)}$ from their respective QCD counterparts as depicted in Fig.~\eqref{Fig:One:loop:Amplitude}$(a)$, $(b)$ and $(d)$, we have the corresponding IR finite hard kernel
%
\bseq
\begin{align}
T_{\mu,\mathrm{em}}^{(1)} (\omega) =& \frac{\alpha_s C_F}{4\pi}T_{\mu}^{(0)}(\omega)\bigg[\left(\ln\frac{q^2+2q^{-}v^{+} \omega}{\mu_F^2}+2\ln\frac{\mu_R}{\mu_F}-4+i\pi\right)
\nn\\
-&\frac{q^{2}}{2q^{-}v^{+} \omega}\ln\frac{q^2+2q^{-}v^{+} \omega}{q^2}\left(3+2\pi^{3}+\textit{i}\pi^{2}\Big\{2\ln\pi+\ln \frac{q^2+2q^{-}v^{+} \omega}{q^2}\Big\}\right)\bigg],
\\
\notag 
T_{\mu,\mathrm{wk}}^{(1)} (\omega)  =& \frac{\alpha_s C_F}{4\pi}T_{\mu}^{(0)}(\omega)\Bigg[2 \ln^2 \frac{z}{\mu _F^2}-\ln^2\frac{q^2+z}{\mu_F^2}-2 \ln\frac{q^2+z}{\mu_F^2}-2\ln^2 \frac{m_b}{\mu _F}+4\ln \frac{m_b}{\mu _F}\Big(1 
\nn\\
+& \ln \frac{q^2+z}{z}+\ln \frac{m_b}{\sqrt{2}q^-}\Big)-2\ln \frac{m_b}{\mu _R}+\Big(2+4 \ln\frac{q^2+z}{z}-2 \ln \frac{\sqrt{2}
q^-}{m_b}\Big) \ln \frac{\sqrt{2} q^-}{m_b}
\nn\\
+&\ln^2 \frac{m_b}{m_b+\sqrt{2}
q^-}+2\text{Li}_2\big(\frac{m_b}{m_b+\sqrt{2} q^-}\big)+\frac{\sqrt{2}  m_b q^-}{q^2+m_b^2+\sqrt{2} m_b q^-} \ln\frac{q^2+\sqrt{2}
m_b q^-}{m_b^2} + \frac{\pi ^2}{12}
\nn\\
-&i \pi\Big\{2 \ln\frac{q^2+z}{\mu_F^2}-2 \ln\frac{\sqrt{2} m_b q^-}{\mu _F^2} + 2 \ln\frac{m_b^2+\sqrt{2} m_b q^-}{\sqrt{2} m_b q^-}+\frac{ \sqrt{2} m_b q^{-}-2q^2}{q^2+m_b^2+\sqrt{2} m_b q^-}\Big\}\Bigg],
\\
T_{\mu,\Sigma}^{(1)}(\omega) = &\frac{\alpha_s C_F}{4\pi}T_{\mu}^{(0)}(\omega)
\left(\ln\frac{q^2+2q^{-}v^{+}\omega}{\mu_R^2}-1+i\pi\right).
\end{align}
\label{T:one:bulk:of:diagrams}
\eseq
%
\hspace{-0.35cm}
where $z=2~q\cdot k$. Now, we look at the wave function correction to the external quark fields. First, we start by considering wave function correction to the $u$-quark, as shown in Fig.~\eqref{Fig:One:loop:Amplitude}$(e)$ and Fig.~\eqref{Fig:One:loop:LCDA}$(e)$.
By employing the LSZ reduction formula, the corresponding hard kernel reads as
\beq
T_{\mu,\delta Z_u}^{(1)}(\omega) =\frac{\alpha_sC_F}{4\pi}\ln\frac{\mu_F}{\mu_R}T_{\mu}^{(0)}(\omega)\,.
\label{T:one-loop:Zu}
\eeq
For external heavy quark field correction, one needs to consider the external $\bar{b}$ correction in QCD amplitude, Fig.~\eqref{Fig:One:loop:Amplitude}$(f)$, and HQET LCDA as illustrated in Fig.~\eqref{Fig:One:loop:LCDA}$(f)$. And then one finds the contribution to $T_{\mu}^{(1)}$ due to the external heavy quark field correction 
\beq
T_{\mu,\delta Z_b}^{(1)} (\omega)=\frac{\alpha_sC_F}{4\pi}
\left(2\ln\frac{m_b}{\mu_F}+\ln\frac{m_b}{\mu_R}-2\right)T_{\mu}^{(0)}(\omega),
\label{T:one-loop:Zb}
\eeq
%
where $\delta Z_q$ represents the standard $q$-quark wave function renormalization constant defined in full QCD. Now we consider the final one-loop graph which is the box diagram as shown in Fig. ~\eqref{Fig:One:loop:Amplitude}$(c)$. At the lowest power of $1/m_b$, one can easily find that only the {\it soft} region of loop momenta ($l^\mu \sim \Lambda_{\rm QCD}$) yields the leading order contribution to $\mathcal{M}^{(1)}$. This contribution is exactly equal to $\Phi^{(1)}_{\mathrm{box}}\otimes T_{\mu}^{(0)}$ and thus a vanishing contribution to $T_{\mu}^{(1)}$ from the box diagram has been found, identical to the case observed in ~\cite{Ishaq:2019zki,DescotesGenon:2002mw}.
Hence, the hard-scattering kernel at NLO in perturbation theory can be obtained by summing up the non-vanishing hard kernel corresponding to each one-loop diagram
\begin{align}
T_{\mu}^{(1)}(\omega) =& \frac{\alpha_s C_F}{4\pi}T_{\mu}^{(0)}(\omega)\Bigg[2 \ln^2 \frac{z}{\mu _F^2}-\ln^2\frac{q^2+z}{\mu_F^2}-2\ln^2 \frac{m_b}{\mu _F}+\ln \frac{m_b}{\mu _F}
\Big(5+4 \ln \frac{q^2+z}{z}+4 \ln \frac{m_b}{\sqrt{2}q^-}\Big)
\nn\\
+&\Big(2+4 \ln\frac{q^2+z}{z}-2 \ln \frac{\sqrt{2}
q^-}{m_b}\Big) \ln \frac{\sqrt{2} q^-}{m_b}+\ln^2 \frac{m_b}{m_b+\sqrt{2}
q^-}+2 \ln\frac{q^2+\sqrt{2} m_b q^-}{\sqrt{2} m_b q^-}
\nn\\
+& 2\text{Li}_2\left(\frac{m_b}{m_b+\sqrt{2} q^-}\right)+\frac{\sqrt{2}  m_b q^-}{q^2+m_b^2+\sqrt{2} m_b q^-}\ln\frac{q^2+\sqrt{2}
m_b q^-}{m_b^2}+\frac{\pi ^2}{12}-7
\nn\\
-&i \pi\Big\{2 \ln\frac{q^2+z}{\mu_F^2}-2 \ln\frac{\sqrt{2} m_b q^-}{\mu _F^2}+2 \ln\frac{m_b^2+\sqrt{2} m_b q^-}{\sqrt{2} m_b q^-}+\frac{2 m_b^2+3 \sqrt{2} m_b q^-}{q^2+m_b^2+\sqrt{2} m_b q^-}\Big\}
\nn\\
+& q^2\Bigg(\frac{\left(2 m_b^2+\sqrt{2} m_b q^-\right)+i \pi  \left(2 m_b^2+3 \sqrt{2} m_b q^-\right)}{\sqrt{2} m_b q^- \left(q^2+m_b^2+\sqrt{2} m_b q^-\right)}\ln\frac{q^2+\sqrt{2} m_b q^-}{m_b^2}
\nn\\
+&\frac{1}{z}\left(3+2 \pi ^3-i \pi ^2 \Big\{2 \ln\pi+\ln\frac{q^2+z}{q^2}\Big\}\right)\ln\frac{q^2+z}{q^2}\Bigg)\Bigg]\,.
\label{T1:full:expression}
\end{align}
The hard-scattering kernel, Eq.~\eqref{T1:full:expression} at NLO precision is IR safe, which guarantees the HQET factorization Eq.~\eqref{HQET:factorization:theorem} is applicable for the exclusive production of heavy-light meson through semileptonic $W$ gauge boson. It is natural to compare Eq.~\eqref{T1:full:expression} with the corresponding result of the rather simplest process, the exclusive production of $B$ meson via radiative $W$ decay~\cite{Ishaq:2019zki}, and found that the expression in Eq.~\eqref{T1:full:expression} at $q^2\rightarrow 0$ reduces to the  corresponding expression for $W^+\to B^+\gamma$~\cite{Ishaq:2019zki}. Now we are in a position to present the central result of this study, the form factors. At leading order expansion in $1/m_b$, the form factor reads
\begin{align}
&F_V^{(1)}  = F_A^{(1)}= {\alpha_s C_F\over 4\pi} f_B m_B\int_0^\infty
\frac{\phi_B^+\big(\omega\big)}{\left(\omega+ q^{+}/v^{+}\right)}
\Bigg[-\ln^2{m_b\over \mu_F} -\ln{m_b\over \mu_F}
\left(2 \ln {1-r\over r}- 2\right)+ 2 {\rm Li}_2 (r)
\nn\\
&-\ln^2(1-r)+2\ln r\ln(1-r)+\left(3-r\right)\ln\frac{1-r}{r}+\frac{\pi^{2}}{12}-5-
2 \ln \frac{\omega+q^+/v^+}{\mu_F} \bigg(\ln\frac{1-r}{r}+\ln\frac{m_b}{\mu_F}\bigg)
\nn\\
&+2 \ln^2\frac{\omega }{\mu_F}-\ln^2 \frac{\omega+q^+/v^+}{\mu_F}+{i \pi}\Big\{2\ln\frac{m_b}{\mu_F}-3+r+2\ln\big(1-r\big)+2 \ln \frac{\omega+q^+/v^+}{\mu_F}\Big\}
\nn\\
&-\frac{q^+}{v^+\omega}\bigg[\bigg(\ln\frac{q^2}{\mu_F^2}-\ln\frac{m_b}{\mu_F}-\ln\frac{1-r}{r}-\ln \frac{\omega+q^+/v^+}{\mu_F}\bigg)
\bigg(3+2\pi^3+i \pi^2 \Big\{\ln\frac{q^2}{\mu_F^2}
\nn\\
&-\ln\frac{m_b}{\mu_F}-\ln\frac{1-r}{r}
-\ln\frac{\omega + q^+/v^+}{\mu_F}
\Big\}
\bigg)
\bigg]
\Bigg]
d\omega\,.
\label{eq:F_VA_01}
\end{align}
where $r\equiv m_b^2/m_W^2$ is a dimensionless constant. At the lowest order in $1/m_b$, the NLO expression of form factors for $W^+\to B^+\ell^+\ell^-$ also depends on the ``+''-momentum of the dilepton pair. This explicit depends on accounting for the difference when compared with the corresponding expression of form factors reported in~\cite{Ishaq:2019zki}. However, in the limit $q^2\rightarrow 0$, the expression in Eq.~\eqref{eq:F_VA_01} coincides with the result presented in~\cite{Ishaq:2019zki}. Here it is important to note that Eq.~\eqref{eq:F_VA_01} exhibits a symmetry relationship among form factors associated with different currents, thereby guaranteeing the preservation of heavy quark spin symmetry at the leading order in the $1/m_b$ expansion.
\section{Numerical Analysis}\label{sec:phnmnlgy}
In this section, we perform numerical computations to predict the vector/axial-vector form factors associated with the $W^+\to B^+\ell^+\ell^-$ process as well as the corresponding decay width and branching fractions. For this purpose, we use the numerical values of input parameters from PDG~\cite{Workman:2022ynf} unless stated otherwise. However, for the evaluation of QCD running coupling $\alpha_s$ at one-loop accuracy, we invoke the atuomated package \texttt{HOPPET}~\cite{Salam:2008qg}

\begin{table}[H]
\begin{centering}
\begin{tabular}{llll}
$\sin\theta_{W}=0.481,\;\;$ & $\alpha\left(m_{W}/2\right)=1/130,\;\;$ & $m_{W}=80.379\;\mathrm{GeV},\;\;$ & $f_{B}=0.190\;\mathrm{GeV},$ \tabularnewline
 $\left|V_{ub}\right|=3.67\times10^{-3},\;\;$ & $m_{b}=4.18\;\mathrm{GeV},\;\;$ & $m_{B}=5.279\;\mathrm{GeV},\;\;$& \tabularnewline
\end{tabular}
\par\end{centering}
\end{table}
It is worth mentioning here that our theoretical predictions are based on the Grozin-Neubert exponential model~\cite{Grozin:1996pq}, where heavy-light meson LCDA at the initial scale $\mu_0=1$ GeV is defined as
\begin{align}
\phi_{B}^{+}\big(\omega\big)=&\frac{\omega}{\lambda_B^2}\exp\left({-\frac{\omega}{\lambda_B}}\right),
\end{align}
with $\lambda_B\equiv \lambda^{+}_B(q^{+}=0)=0.350\pm0.15$ GeV~\cite{Gelb:2018end} .

\begin{figure}[!h]
\centering\scalebox{1}{
\begin{tabular}{ccc}
\includegraphics[width=4in,height=3in]{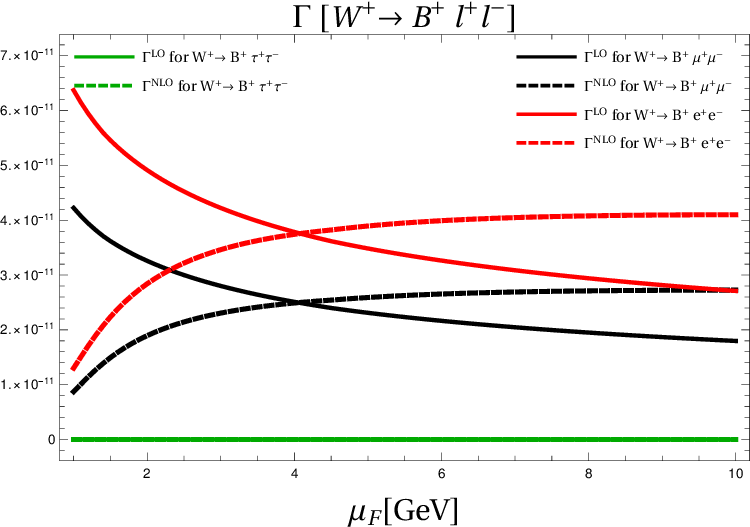}
\end{tabular}}
\caption{Decay rates of $W^+\to B^+\ell^+\ell^-$ as a function of $\mu_F$, which varies from $1$ to $10$ GeV.}
\label{BRVsFactScale}
\end{figure}

In the current study, we have calculated the leading order (LO) and next-to-leading order (NLO) QCD corrections to the form factors for the decay $W^+\to B^+\ell^+\ell^-$ which are defined as $F_{V/A}^{\text{LO}}\equiv F_{V/A}^{(0)}$ and $F_{V/A}^{\text{NLO}}\equiv F_{V/A}^{(0)}+ F_{V/A}^{(1)}$. The form factors $F_{V/A}^{(0)}$ and $ F_{V/A}^{(1)}$ are given in Eq. \eqref{FVA:LO:expression} and Eq.\eqref{eq:F_VA_01}, respectively. One can see from these equations that the form factors rely on the $B$ meson LCDA, which demonstrates scale dependence $\mu_F$. Therefore, to understand the variation in the vector/axial-vector form factors arising due to the factorization scale $\mu_F$, it is essential to first know the $\mu_F$ dependence of LCDA. Consequently, one can calculate the sensitivity of physical observables such as decay rates and branching fractions to the scale $\mu_F$. To achieve this goal, we use the analytical solutions of Lange-Neubert evolution Eq. \eqref{LN:evolution:eq} as reported in ~\cite{Lee:2005gza,Bell:2013tfa}, to obtain the form factors at desired scales as a function of the invariant squared-mass of the dilepton, $q^2$.
\begin{figure}[h]
\centering\scalebox{1}{
\begin{tabular}{ccc}
\includegraphics[width=3in,height=2in]{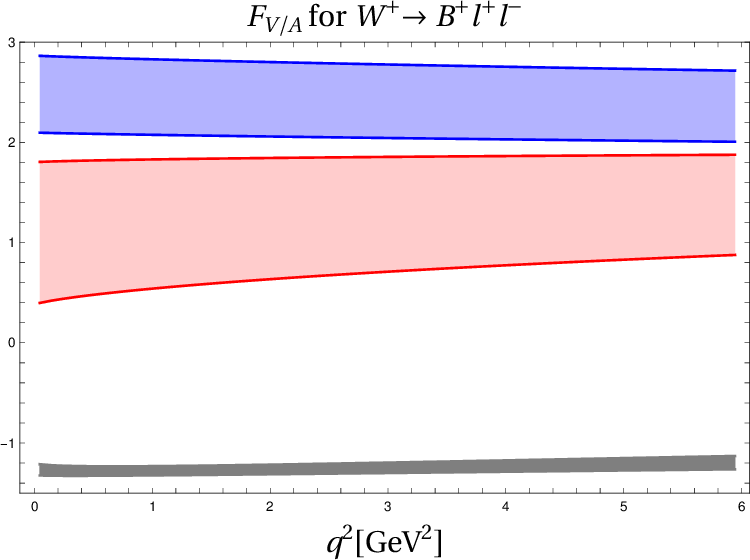} &
\includegraphics[width=3in,height=2in]{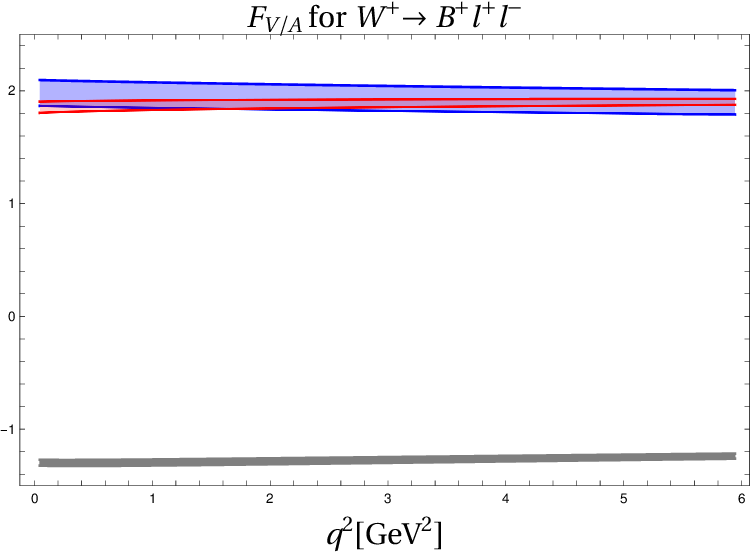}\\
    (a)    &    (b)\\
\end{tabular}}
\caption{The $q^2$ dependence of the vector/axial vector form-factors at LO and NLO in $\alpha_s$. The band represents the uncertainty from $\mu_F=1$ to $m_B$ (left) and $\mu_F=m_B$ to $10$ GeV (right).}
\label{FFVsq2}
\end{figure}
In Fig. \eqref{FFVsq2}, we plot the form-factors at LO and NLO in $\alpha_s$ as a function of $q^2$. The band indicates the uncertainty arising from the factorization scale, $\mu_F$. We divide the variation in $\mu_F$ into two intervals: $1\,\GeV\le\mu_1\le m_B$ (left) $\&$ $m_B\le\mu_2\le 10\, \GeV$ (right), one can notice that the $q^2$ dependence of the form factors is very mild for both intervals while the color bands depict the scale dependence. The blue, red, and gray bands correspond to the LO form factors ($F_{V/A}^{\text{NLO}}$), the real part of NLO form factors ($\mathcal{R}e[F_{V/A}^{\text{NLO}}]$) and the imaginary part of the NLO form factors ($\mathcal{I}m[F_{V/A}^{\text{NLO}}]$), respectively. It is observed that these uncertainty bands of the $F_{V/A}^{\text{LO}}$ and $F_{V/A}^{\text{NLO}}$ (both computed at percision in $\alpha_s$) turn out to be well separated as $\mu$ varies within the $\mu_1$ interval. While for the $\mu_2$-interval, the uncertainty bands reduce and overlap for both $F_{V/A}^{\text{LO}}$ and $\mathcal{R}e[F_{V/A}^{\text{NLO}}]$, however, the $\mathcal{I}m[F_{V/A}^{\text{NLO}}]$ is not significantly changed. As one can also see from Fig. \eqref{FFVsq2}(b) that the reduction in the $F_{V/A}^{\text{NLO}}$ is greater than the $F_{V/A}^{\text{LO}}$ for $m_B\le\mu_2\le 10\, \GeV$ which is attributed to the inclusion of $\alpha_{s}$ correction in the form-factors. This ensures the decay rates at NLO for $W^+\to B^+\ell^+\ell^-$ processes in the $\mu_2$-interval are almost insensitive as shown in Fig.\eqref{BRVsFactScale}. However, for relatively small scales, the NLO form factors, $F_{V/A}^{\text{NLO}}$, still reflect the notable dependence on $\mu_F$. This residual scale dependence could be eliminated by incorporating the higher-order QCD corrections.\\
Moreover, in our NLO predictions for decay rates, we also include the imaginary part of one loop corrections to the form factors, $\mathcal{I}m[F_{V/A}^{(1)}]$, without strictly truncating the decay width at $\mathcal{O}(\alpha_s)$. To show the $\mu_F$-dependence, for the decay rates $W^+\to B^+\ell^+\ell^-$ ($\ell=e,\mu,\tau$) against $\mu_F$, we integrated over $q^2\epsilon[4m_\mu^2,6]$ for muon and to avoid the photon pole we use the same $q^2$-bin for electron as well, while for tauon we take $q^2\epsilon[14,20]$ and plotted in Fig. \eqref{BRVsFactScale}. From this figure, one can see that for the case of tauons as a final state leptons, the decay rate is around three order suppressed in comparison of electron and muon cases which is quantified in Tab. \ref{Tab.nmrcl_prdctn}. Where as for $W^+\to B^+\ell^+\ell^-$ ($\ell=e,\mu$), the LO decay rates ($\Gamma^{\mathrm{LO}}$) strongly depend upon the $\mu_F$. Similarly, the NLO decay rates ($\Gamma^{\mathrm{NLO}}$) at a low $\mu_F$ scale, from 1 GeV to 4 GeV, are also highly sensitive. On the other hand, the sensitivity is very mild at a relatively large scale, above than 4 GeV. This feature emerges because one-loop QCD corrections generate significant precision in the form factors. Consequently, the scale dependence in the NLO decay rates ($\Gamma^{\mathrm{NLO}}$) gets largely reduced, particularly for relatively large $\mu_F$. The profiles of the decay rates against the factorization scale show similar behavior as seen in $W^+\to B^+\gamma$~\cite{Ishaq:2019zki}. We have also calculated the numerical values of these decay rates ($\Gamma^{\mathrm{LO}}$, $\Gamma^{\mathrm{NLO}}$) and the branching fractions ($\mathrm{Br}^{\mathrm{NLO}}$) at NLO by varying the $\mu_F$ from 1 Gev to 10 GeV, which are listed in Tab. \ref{Tab.nmrcl_prdctn}. We found that the NLO corrections turn out to be substantial, which may vary from $-76\% \text{ to } +58\%$ of the LO decay rates for the case of electron and $-79\% \text{ to } +52\%$ for muon, whereas for tauon it varies from $-63\% \text{ to } +72\%$.
\begin{table}[H]
\begin{centering}
\scalebox{0.88}{
\begin{tabular}{|c|c|c|c|}
\hline
Decay Channel & $\vphantom{\frac{L^L}{L^L}}\Gamma^{\mathrm{LO}}$ & $\Gamma^{\mathrm{NLO}}$ & $\mathrm{Br}^{\mathrm{NLO}}$\tabularnewline
\hline
$\vphantom{\frac{L^L}{L^L}}W^{+}\rightarrow B^{+}e^+e^-$ & $\left(6.37-2.71\right)\times10^{-11}$GeV & $\left(1.51-4.10\right)\times10^{-11}$GeV & $\left(0.72-1.97\right)\times10^{-11}$\tabularnewline
$\vphantom{\frac{L^L}{L^L}}W^{+}\rightarrow B^{+}\mu^+\mu^-$ & $\left(4.23-1.80\right)\times10^{-11}$GeV & $\left(0.87-2.73\right)\times10^{-11}$GeV & $\left(0.42-1.31\right)\times10^{-11}$\tabularnewline$
\vphantom{\frac{L^L}{L^L}}W^{+}\rightarrow B^{+}\tau^+\tau^-$ & $\left(2.39-1.06\right)\times10^{-14}$GeV & $\left(0.87-1.82\right)\times10^{-14}$GeV & $\left(0.42-0.87\right)\times10^{-14}$\tabularnewline
\hline
\end{tabular}}
\par\end{centering}
\caption{Numerical predictions to the decay rates and branching ratios for the processes 
$W^+\to B^+\ell^+\ell^-$ with $\ell=e,\mu,\tau$ . The uncertainty is estimated by varying $\mu_F$ from 1 GeV to 10 GeV at $\lambda_B=0.35$ GeV after integrating over $q^2\epsilon[4m_\mu^2,6]$ for electron and muon while for tauon integrating over $q^2\epsilon[14,20]$.}
\label{Tab.nmrcl_prdctn}
\end{table}
\begin{figure}[!h]
\centering\scalebox{1}{
\begin{tabular}{ccc}
\includegraphics[width=3in,height=2in]{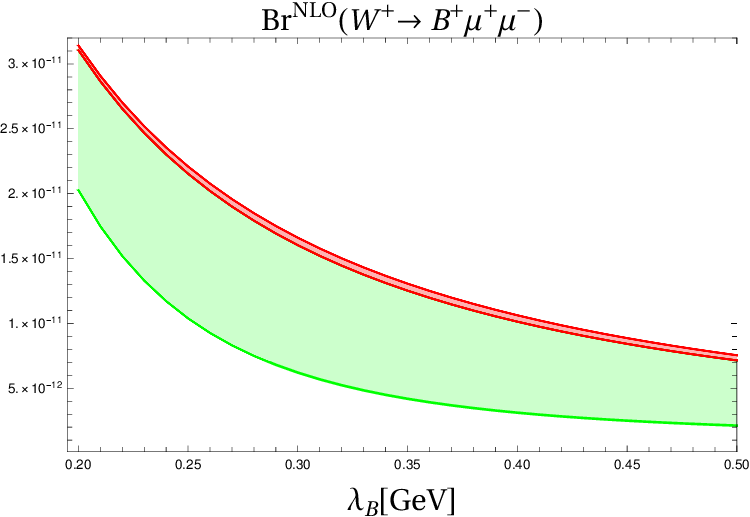} &
\includegraphics[width=3in,height=2in]{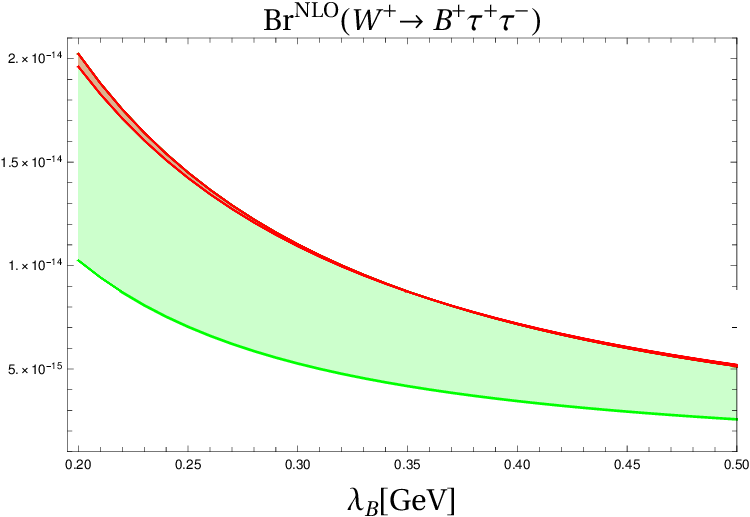}\\
    (a)    &    (b)\\
\end{tabular}}
\caption{Illustration of the inverse moment $\lambda_{B}(\mu_0)$ dependence of the branching fraction for the decay modes $W^+\to B^+\ell^+\ell^-$. The uncertain band shows the variation in $\mu_F$ from $1$ GeV to meson mass.}
\label{BRVsOmega}
\end{figure}
In addition, the theoretical predictions for the branching ratios are also influenced by the parameter $\lambda_B(\mu_0)$, because it affects the HQET factorization through $\phi_{B}^{+}$. Therefore, the analysis of branching fractions as a function of $\lambda_B$ is a handy tool to precisely constraint this parameter.
For this purpose, to see the sensitivity of the branching fraction to the $\lambda_B$, we plotted it against the $\lambda_B$ by using the range: $\lambda_B(\mu_0)=0.35\, \GeV \pm 0.15 \, \GeV $~\cite{Gelb:2018end} and
shown in Fig.~\eqref{BRVsOmega} by the green and red bands. Fig.~\eqref{BRVsOmega}(a) depicts when the leptons in the final state are muons, while Fig.~\eqref{BRVsOmega}(b) corresponds to the case of tauon as a final state leptons. The width of the green and red bands represents the variation by $\mu_F$ in the interval $1\,\GeV\le\mu_1\le m_B$ and $m_B\le\mu_2\le 10\, \GeV$, respectively.

The green band indicates a strong dependence on $\lambda_B$ compared to uncertainty arising from the scale $\mu_F$. Similarly, the branching fraction exhibits higher sensitivity to the smaller values of $\lambda_B$ compared to its larger values. Therefore, the branching fraction in the interval $m_B\le\mu_2\le 10\, \GeV$ is more suitable to extract the precise value of $\lambda_B$, particularly around the lower value of $\lambda_B\simeq0.24$ GeV, which is measured by Belle with $90\%$ C.L.

To further explore how the parametric dependence of the decay rates vary in the different $q^2$ bins, the numerical values of the decay rates are calculated for the processes $W\to B^+\ell^+\ell^-$ where $\ell=e,\mu,\tau$ and listed in Tab. \ref{wc table}. To get the numerical values, we have integrated over three $q^2$ bins: $[4m_\mu^2,0.96]$, $[4m_{\mu}^2,6]$ and $[2,6]$ for the case of electron and muon, while for tauon, we have selected the $q^2$ bin above than the $c\bar c$ resonance region, i.e., $[14,20]$. In the first and second columns, we have listed the numerical values for the hard ($\mu_h$) and hard-collinear ($\mu_{hc}$) scales by setting the ${\mu_{hc}=1.5,\mu_h=5}$ GeV and $\lambda_B=0.35$ GeV for the LO and NLO, respectively. In the remaining columns, we have given the uncertainties in the decay rates by using the ranges of parameters: $\mu_{hc}=1.5\pm0.5$ GeV, $\mu_h=5^{+5}_{-2.5}$ GeV, and $\lambda_B=0.35\pm0.15$ GeV. The total uncertainty is calculated by adding the uncertainties due to the $\mu_{h, hc}$ and $\lambda_B$ in quadrature.
\begin{table*}[!h]
\scalebox{0.6}{
\begin{tabular}{c|c|c|c|c|c|c|c|c|c}
\hline\hline
\multirow{3}{*}{Decay}&q$^2$ bin&LO&NLO&\multicolumn{6}{c}{Uncertainty}\\ 
 &  GeV$^2$  & $\lambda_B$=0.35 GeV  &   $\lambda_B$=0.35 GeV  & LO & NLO & LO $(\lambda_B$) & NLO $(\lambda_B$) &  LO (tot) & NLO (tot) \\ 
 &    & ($\mu_{hc},\mu_h$)=(1.5,5)GeV   &  ($\mu_{hc},\mu_h$)=(1.5,5)GeV   & ($\mu_{hc},\mu_h$) & ($\mu_{hc},\mu_h$) & (1.5,5)GeV & (1.5,5)GeV & (1.5,5)GeV & (1.5,5)GeV \\ \hline
 \ \  \multirow{3}{*}{$B^+ e^+e^-$} \ \  & \ \  $[4m_\mu^2,0.96]$ \ \  & $(5.25,3.35)\times10^{-11}$  & $(2.14,3.74)\times10^{-11}$  &  $(^{+0.88}_{-0.52},^{+1.00}_{-0.75})$ & $(^{+0.58}_{-0.90},^{+0.20}_{-0.65})$ & $(^{+9.72}_{-2.56},^{+0.49}_{-1.47})$ & $(^{+4.27}_{-0.93},^{+5.56}_{-1.60})$ & ($^{+9.76}_{-2.61},^{+1.11}_{-1.65}$) & $(^{+4.31}_{-1.29},^{+5.56}_{-1.73})$ \\ 
 &  $[4m_\mu^2,6]$  & $(5.46,3.48)\times10^{-11}$  & $(2.24,3.90)\times10^{-11}$  &  $(^{+0.91}_{-0.54},^{+1.04}_{-0.77})$ & $(^{+0.61}_{-0.93},^{+0.21}_{-0.68})$ & $(^{+10.09}_{-2.66},^{+5.14}_{-1.53})$ & $(^{+4.45}_{-0.97},^{+5.79}_{-1.67})$ & $(^{+10.13}_{-2.71},^{+5.24}_{-1.71})$ & $(^{+4.49}_{-1.34},^{+5.79}_{-1.80})$ \\ 
 & $[2,6]$  &  $(7.78,5.00)\times10^{-13}$ & $(3.81,6.01)\times10^{-13}$ & $(^{+1.27}_{-0.76},^{+1.47}_{-1.10})$ & $(^{+0.86}_{-1.52},^{+0.19}_{-0.84})$ & $(^{+13.71}_{-3.74},^{+7.04}_{-2.16})$ & $(^{+6.77}_{-1.65},^{+8.72}_{-2.57})$ & ($^{+13.77}_{-3.82},^{+7.19}_{-2.42}$) & $(^{+6.82}_{-2.24},^{+8.72}_{-2.70})$ \\ \hline
 \ \  \multirow{3}{*}{$B^+ \mu^+\mu^- $} \ \  & \ \  $[4m_\mu^2,0.96]$ \ \  & $(3.41,2.18)\times10^{-11}$
 & $(1.40,2.44)\times10^{-11}$  &  ($^{+0.57}_{-0.34},^{+0.65}_{-0.48}$) & $(^{+0.38}_{-0.58},^{+1.29}_{-0.42})$ & ($^{+6.31}_{-1.66},^{+3.22}_{-0.96}$) & $(^{+2.79}_{-0.61},^{+3.62}_{-1.05})$ & ($^{+6.34}_{-1.69},^{+3.29}_{-1.07}$) & $(^{+2.82}_{-0.84},^{+3.84}_{-1.13})$ \\ 
 &  $[4m_\mu^2,6]$  & $(3.62,2.31)\times10^{-11}$  & $(1.49,2.59)\times10^{-11}$  &  $(^{+0.61}_{-0.36},^{+0.69}_{-0.51})$ & $(^{+0.40}_{-0.62},^{+0.14}_{-0.45})$ & ($^{+6.68}_{-1.76},^{+3.40}_{-1.01}$) & $(^{+2.97}_{-0.65},^{+3.85}_{-1.11})$ & ($^{+6.71}_{-1.80},^{+3.47}_{-1.13}$) & $(^{+3.00}_{-0.90},^{+3.85}_{-1.20})$ \\ 
 & $[2,6]$  &  $(7.72,4.97)\times10^{-13}$ & $(3.78,5.96)\times10^{-13}$ &  $(^{+1.26}_{-0.76},^{+1.45}_{-1.09})$ & $(^{+6.72}_{-1.64},^{+8.66}_{-2.55})$ & ($^{+13.60}_{-3.71},^{+6.99}_{-2.15}$)& $(^{+2.42}_{-0.60},^{+3.13}_{-0.93})$ & ($^{+13.66}_{-3.79},^{+7.14}_{-2.41}$) & $(^{+7.14}_{-1.75},^{+9.21}_{-2.71})$ \\ \hline
  \ \ $B^+\tau^+\tau^-$ \ \  & $[14,20]$  & ($0.21,0.14)\times10^{-13}$  & $(0.13,0.18)\times10^{-13}$  & $(^{+0.32}_{-0.19},^{+0.38}_{-0.29})$ & $(^{+0.45}_{-0.22},^{+0.00}_{-0.16})$ & ($^{+3.23}_{-0.95},^{+1.70}_{-0.56}$) & ($^{+1.89}_{-0.55},^{+2.40}_{-0.75}$) & ($^{+3.25}_{-0.97},^{+1.74}_{-0.63}$) & ($^{+1.94}_{-0.59},^{+2.4}_{-0.77}$) \\
\hline\hline
\end{tabular}}
\caption{The numerical values of decay rates are integrated over the different $q^2$ bins for the processes $W\to B^+\ell^+\ell^-$ where $\ell=e,\mu,\tau$. In the first and second columns, we have listed the numerical values by setting the ${\mu_{hc}=1.5\, \GeV,\mu_h=5}$ GeV and $\lambda_B=0.35$ GeV for the LO and NLO, respectively. In the remaining columns, we have given the uncertainties in the decay rates by using the ranges of parameters: $\mu_{hc}=1.5\pm0.5$GeV, $\mu_h=5^{+5}_{-2.5}$ GeV, and $\lambda_B=0.35\pm0.15$ GeV. The total uncertainty is calculated by adding the uncertainties due to the $\mu_{h, hc}$ and $\lambda_B$ in quadrature. $\lambda_B$ in quadrature.} \label{wc table}
\end{table*}
\section{Summary}\label{sec:sumry}
In this paper, we have studied the production of heavy-light meson at NLO in strong coupling $\alpha_s$ at leading order in $1/m_b$ through semileptonic $W$ boson decay in the HQET factorization approach. Here, we have extended the HQET factorization formula for a rather sophisticated decay for $W\to B\ell^+\ell^-$ validating the scale hierarchy, $m_W\sim m_b\gg \LmbdQCDD$, in a similar fashion to that reported in ~\cite{Ishaq:2019zki}. However, the amplitude of the decay process under consideration is highly suppressed when $q^2\sim\mathcal{O}(m_B^2)$. However, we have shown that the amplitude for the exclusive production of a heavy-light meson can be factorized at both the soft scale, $q^2 \ll m_b^2$, and the intermediate scale, $q^2\sim\mathcal{O}(m_b\,\LmbdQCDD)$. To achieve this goal, we explicitly calculated the hard scattering kernel at NLO, which is IR finite.

In addition, we have calculated the form factors associated with the process $W\to B\ell^+\ell^-$ up to $\alpha_s$ at the lowest order in $1/m_b$. It has been determined that both of these form factors are identical up to $\alpha_s$ at the lowest order in $1/m_b$. This confirms the heavy quark spin symmetry and supports the accuracy of our results. Furhtermore, it is important to emphasize that our results for form factors and hard kernels are consistent with the corresponding findings of Ref.~\cite{Ishaq:2019zki} when the value of $q^2$ approaches zero. We have also investigated the impact of NLO perturbative corrections for the numerical prediction of vector and axial-vector form factors appearing in $W^+\to B^+\ell^+\ell^-$ and presented the $q^2$ as well as factorization scale $\mu_F$ dependence of these form factors. The results of our study suggest that the form factors show very little variation for $q^2$. Additionally, within the range of $m_B\le\mu_F\le 10\, \GeV$, the form factors also become unaffected by varies in the factorization scale $\mu_F$. 

Moreover, by using these form factors, we have calculated the decay rates and branching ratios for the processes $W\to B^+\ell^+\ell^-$ where $\ell=e,\mu,\tau$ at $\lambda_B=0.35$ GeV. We have observed significant NLO corrections, ranging from a decrease of $-76\%$ to an increase of $+58\%$ in the decay rates for electrons; for muons the range is $-79\%$ to $+52\%$. Although for the case of tauons the decay rate is suppressed by around three orders in comparison with electron and muon cases, the NLO corrections range from a decrease of $-63\%$ to an increase of $+72\%$, depending on the variation of the scale parameter $\mu_F$ from 1 - 10 GeV. The factorization formula in the current study is valid for both low- and intermediate-$q^2$ regions. Therefore, we have also computed the numerical values of the decay rates in different $q^2$ bins. It is found that the branching ratios are sensitive to the first inverse moment $\lambda_B$, particularly for relatively large values of the factorization scale. Therefore, the branching ratios of $W\to B^+\ell^+\ell^-$ within the range of $m_B\le\mu_F\le 10\, \GeV$ are more suited to constraint the value of $\lambda_B$. Hence, this decay channel could be used as an additional source to pin down the precise value of $\lambda_B$, which explicitly appears in both the production and decay processes of $B$ mesons.


\end{document}